\documentclass[traditabstract,twocolumn]{aa} 

\usepackage{epsfig,graphicx,natbib,amsmath}

\begin{document}

\title{CO rotational line emission from a dense knot in Cas~A}
\subtitle{Evidence for active post-reverse-shock chemistry}

\author{Sofia H. J. Wallstr\"om \inst{1}
\and Chiara Biscaro \inst{2}
\and Francisco Salgado \inst{3}
\and John H. Black \inst{1}
\and Isabelle Cherchneff \inst{2}
\and S\'ebastien~Muller \inst{4}
\and Olivier Bern\'e \inst{5,6}
\and Jeonghee Rho \inst{7,8}
\and Alexander G. G. M. Tielens \inst{3}
}

\institute{Department of Earth and Space Sciences, Chalmers University of Technology, SE-43992 Onsala, Sweden
\and Department Physik, Universit\"at Basel, CH-4056 Basel, Switzerland
\and Leiden Observatory, Leiden University, P.O. Box 9513, NL-2300 RA, The Netherlands
\and Onsala Space Observatory, Chalmers University of Technology, SE-43992 Onsala, Sweden
\and Universit\'e de Toulouse, UPS-OMP, IRAP, 31028 Toulouse, France
\and CNRS; IRAP; 9 Av. Colonel Roche, BP 44346, F-31028 Toulouse cedex 4, France
\and SETI Institute, 189 N. Bernardo Ave, Mountain View, CA 94043
\and Stratospheric Observatory for Infrared Astronomy, NASA Ames Research Center, MS 211-3, Moffett Field, CA 94035, USA
}

\date {Received  / Accepted}

\titlerunning{CO in Cas~A}
\authorrunning{Wallstr\"om et al.}

\abstract{We report a {\it Herschel}\thanks{{\it Herschel} is an ESA space observatory with science instruments provided by European-led Principal Investigator consortia and with important participation from NASA.} detection of high-J rotational CO lines from a dense knot in the supernova remnant Cas~A. Based on a combined analysis of these rotational lines, and previously observed ro-vibrational CO lines, we find the gas to be warm (two components at $\sim$400 and 2000~K) and dense (10$^{6\text{-}7}$ cm$^{-3}$), with a CO column density of $\sim$5$\times$10$^{17}$~cm$^{-2}$. This, along with the broad line widths ($\sim$400 km\,s$^{-1}$), suggests that the CO emission originates in the post-shock region of the reverse shock. As the passage of the reverse shock dissociates any existing molecules, the CO has most likely reformed in the last few years, in the post-shock gas. The CO cooling time is comparable to the CO formation time, so possible heating sources (UV photons from the shock front, X-rays, electron conduction) to maintain the large column density of warm CO are discussed.

}

\maketitle

\section{Introduction}\label{intro}

Stars with masses ranging from 8 to 30 M$_\odot$ have lifetimes of some 10$^7$ years \citep{woo02} before exploding as Type II supernovae (SNe), enriching their local environments on a short timescale. Dust grains and molecules are produced in the ejected material (ejecta), despite the harsh physical conditions. Emission from CO and SiO molecules has been observed at infrared (IR) wavelengths in SN1987A some hundred days after the SN explosion \citep{dan87,luc89,roc91}, and in several other SNe \citep{kot05,kot06,kot09}. These observations demonstrate that a rapid and efficient chemistry develops in the ejecta, and that molecular formation is a common occurrence in SNe \citep{lep90,che11}.
CO and SiO have been observed in the ejecta of SN1987A \citep{kam13}, demonstrating that these molecules have survived up to 25 years after the SN explosion.

SNe are prime contenders for explaining the large dust masses in the early universe, inferred from the reddening of background quasars and Lyman-$\alpha$ systems at high redshift \citep{pei91,pet94,ber03}, as efficient dust formation on short timescales is required. However, IR observations of SNe $\sim$400 days post-explosion show dust masses ranging between $10^{-5}$ and $10^{-2}$ M$_\odot$ (e.g., \citealp{luc89,sug06,sza13}), at least an order of magnitude too little to explain the observed high-z dust \citep{dwe08}. Larger dust masses have been observed in supernova remnants (SNRs), for example $\sim$0.7 M$_\odot$ of cool dust in the young remnant of SN1987A \citep{mat11}. Observations of the SNR Cas~A imply $\sim$0.025 M$_\odot$ of warm dust \citep{rho08}, and $\sim$0.075 M$_\odot$ of cool  dust \citep{bar10}. However, these dust masses do not necessarily represent the net SN dust yields into the ISM, as the passage of a reverse shock might reprocess the dust grains. 

When a SN explosion shock wave has swept up enough circumstellar and interstellar matter, a reverse shock forms and travels inwards, decelerating and reprocessing the ejecta \citep{che77}. At the shock front the ejecta will be heated to $>$10$^6$~K, sputtering dust and dissociating molecules. However, the shock can be attenuated in dense knots in the ejecta, mitigating its destructive effects there. 

The 330-year-old SNR Cas~A (D=3.4 kpc) is the perfect laboratory for studying the effects of the reverse shock, as it is just beginning to reprocess the ejecta. Ro-vibrational CO emission has been detected in Cas~A \citep{rho09,rho12}, in several ($\sim$20) small ($<$0.8$\arcsec$) knots coincident with the reverse shock. Such knots are reminiscent of the fast-moving knots (FMKs) seen in the optical \citep{fes01}. 
In order to determine the chemical and physical conditions in these knots, we used the Herschel PACS instrument to observe several high-J rotational CO lines towards the brightest CO knot in the remnant.

\section{Observations}\label{obs}

\begin{figure}[t] 
\includegraphics[width=9cm]{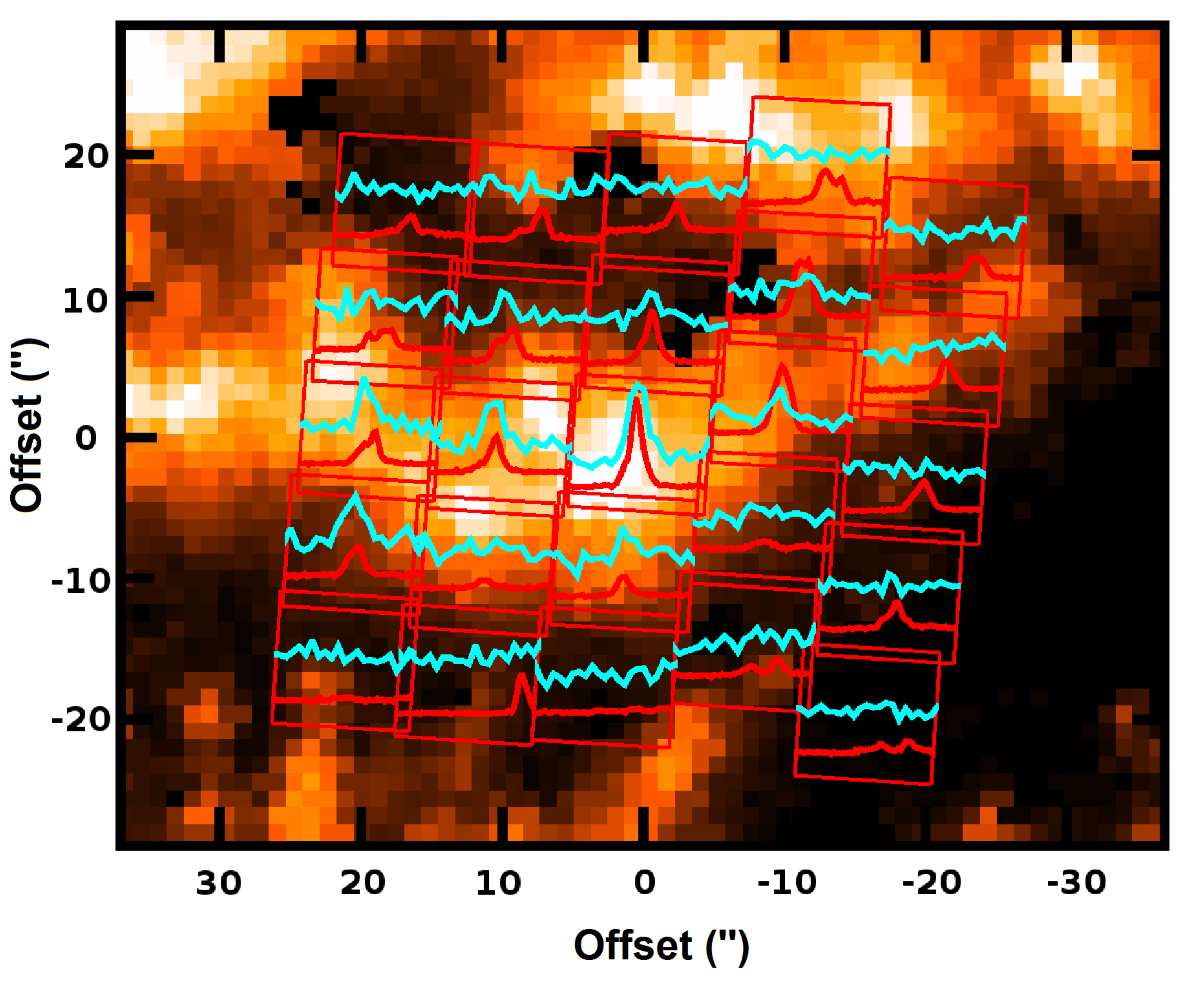} 
\caption{\label{fig:footprint} The PACS footprint, centered on RA 23:23:24.9 Dec +58:50:03.3, showing the spectra of CO J=23-22 in blue and [$\ion{O}{III}$] 88$\mu$m in red, overlaid on a Spitzer/IRAC image of CO vibrational emission in Cas A.}
\end{figure}

The Photodetector Array Camera and Spectrometer (PACS; \citealp{pog10}) aboard the Herschel Space Observatory \citep{pil10} was used on 2012 June 11, to observe a single pointing in the northwest of Cas~A (Fig.~\ref{fig:footprint}). PACS consists of a 5$\times$5 array of spatial pixels (spaxels), each 9.4$\arcsec$$\times$9.4$\arcsec$ in size.
We selected seven CO rotational transitions, between J$_{up}$=14 and 38, for even coverage of the CO ladder in the PACS spectral range. The observations were done using short range spectroscopy scans with high spectral resolution (R=1000--2700, $\sim$100--300 km\,s$^{-1}$).
The [$\ion{O}{III}$] 88 $\mu$m transition was observed within the covered spectral range. 
Observations were done in chopping/nodding mode with a chop throw of 6$\arcmin$ to the north and south, well off the remnant. 

The data were processed using the standard range scan reduction pipeline, implemented in HIPE version 9.0.0 \citep{ott10}. The CO J=38--37 spectrum was marred by an artifact and thus discarded. A first degree polynomial baseline was fit from line-free channels and subtracted. The final spectra correspond to the central spaxel after applying the standard PACS point source correction, to correct for flux spill-over into adjacent spaxels. PACS has an absolute flux accuracy of $\sim$10\% (PACS Observer's Manual).

\begin{figure}[t] \begin{center}
\includegraphics[width=\linewidth]{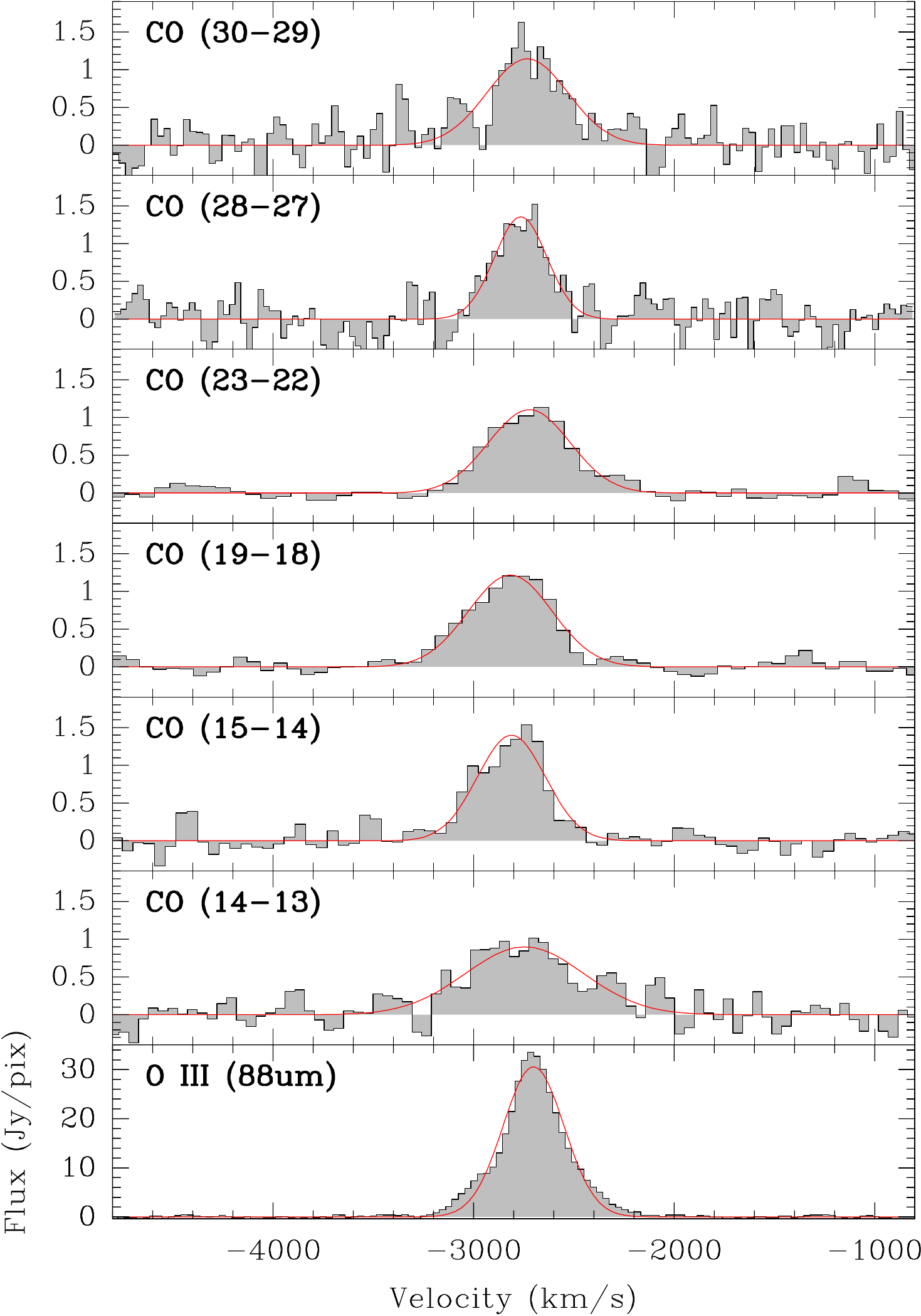}
\caption{\label{fig:spec} CO, and $\ion{O}{III}$, emission lines in velocity (LSR), with Gaussian fits, extracted from the central PACS spaxel. Note that the flux density scale is different for the $\ion{O}{III}$ line. }
\end{center} \end{figure}

\section{Results}\label{results}

All the targeted CO lines are detected in the central PACS spaxel. The line profiles extracted from the central spaxel (Fig.~\ref{fig:spec}) were fit with gaussians (Table~\ref{tab:cenpixdat}). 
The lines have an average radial velocity of --2740~km\,s$^{-1}$, consistent with the velocity of the IR knot n2 \citep[close to our target position]{rho12}, and are broad, with a deconvolved FWHM $\sim$400~km\,s$^{-1}$. 

\begin{figure}[t] \begin{center}
\includegraphics[width=6cm,angle=270]{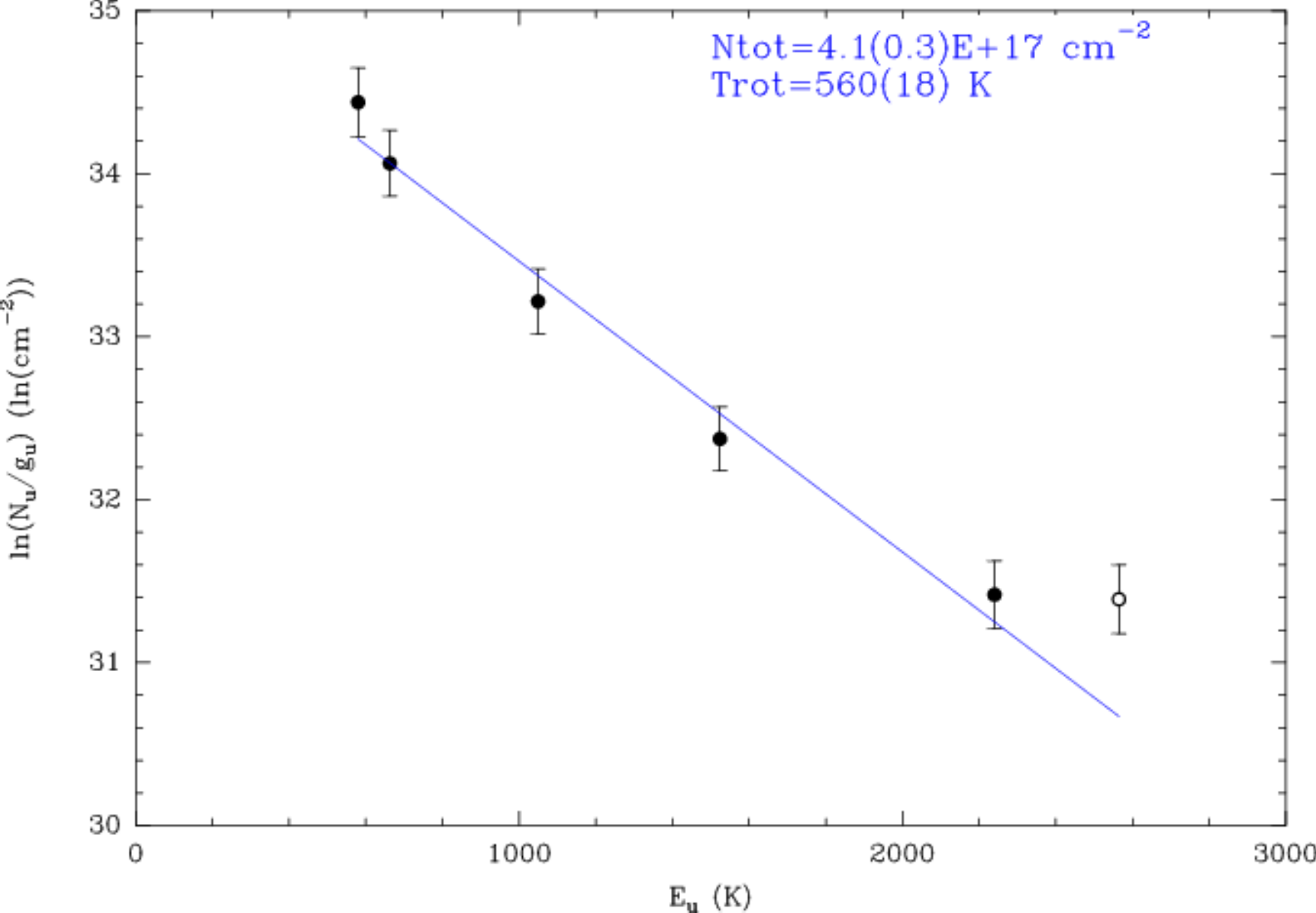}  
\caption{\label{fig:exdi} CO excitation diagram, plotting ln($N_u/g_u$) vs. $E_u$. The column density, N$_{tot}$, and rotation temperature, T$_{rot}$, and their uncertainties, were calculated by Monte Carlo method, excluding the last point; the blue line corresponds to those best-fit parameters.}
\end{center} \end{figure}

The integrated CO line fluxes were used to create a rotational diagram \citep{gol99}, as shown in Fig.~\ref{fig:exdi}.
We adopt a source diameter of 0.5$\arcsec$ (much smaller than the spaxel size), consistent with the 0.8$\arcsec$ upper limit on the near-IR ro-vibrational CO source size \citep{rho09} and optical observations of FMKs \citep{fes01}.
Assuming local thermal equilibrium (LTE) conditions, we derive a column density of $N_{CO}$=(4.1$\pm$0.3)$\times$10$^{17}$~cm$^{-2}$ and a rotation temperature of $T_{rot}$=560$\pm$20~K. 
This corresponds to a CO mass of $\sim$5$\times10^{-6}$~M$_\odot$ for the knot, which is independent of the assumed knot size, as $N_{CO}$ is inversely proportional to the knot size squared.

In the spectra of the full PACS footprint (Fig.~\ref{fig:footprint}) we see that the central spaxel emission has spilled over into the adjacent spaxels, because the point spread function is larger than the spaxel size. There is also some additional CO emission at the eastern edge of the footprint. This is most likely indicative of a second knot, rather than extended CO emission, given the IR and optical evidence for a multitude of small, dense knots. Due to the increased uncertainty at the edge of the PACS footprint, and the smaller CO fluxes, we do not attempt to analyze this emission. 
In contrast to the CO, the emission of the [$\ion{O}{III}$] line is extended over the PACS footprint, which might be explained by its lower critical density as compared with the high-J CO lines.

\begin{table}[t] \begin{center}
\caption{\label{tab:cenpixdat} Data on the CO and [$\ion{O}{III}$] lines extracted from the central spaxel. v$_0$ is the radial velocity of the line, and $\delta$v the velocity full width at half maximum. Errors are given in parentheses.}
\begin{tabular}{lcccc}
\hline \hline
Line & $\lambda$ & v$_0$ & $\delta$v & $\int$Flux\,\textit{dv} \\
 & $\mu$m & km\,s$^{-1}$ & km\,s$^{-1}$ & 10$^{-18}$ W\,m$^{-2}$ \\
\hline
CO (14-13) & 186.00 & -2882(370) & 661(63) & 35(2) \\
CO (15-14) & 173.63 & -2814(362) & 336(24) & 34(2) \\
CO (19-18) & 137.20 & -2814(336) & 409(19) & 46(2) \\
CO (23-22) & 113.46 & -2716(309) & 372(19) & 50(2) \\
CO (28-27) & 93.35 & -2765(105) & 292(32) & 48(4) \\
CO (30-29) & 87.19 & -2733(470) & 418(46) & 64(6) \\
{[}$\ion{O}{III}$] & 88.36 & -2701(100) & 341(7) & 1310(20) \\
\hline \hline
\end{tabular}
\end{center} \end{table}

\begin{figure}[t] \begin{center}
\includegraphics[width=10cm]{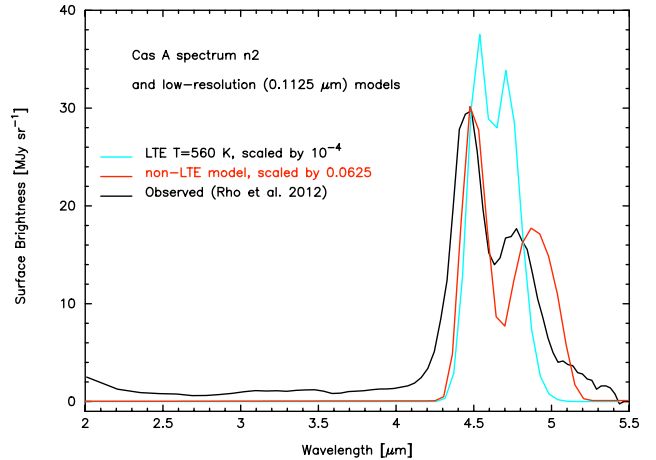} 
\caption{\label{fig:AKARI} The mid-IR AKARI spectrum reported by \citep{rho12} for knot n2 in Cas A is compared with models that have been convolved to the same resolution, 0.1125 $\mu$m. The black curve is the observed spectrum. The red curve is the computed spectrum for the same two-component non-LTE model that best describes the fluxes of the CO rotational lines, but scaled in intensity by a factor 1/16. The cyan curve represents the vibrational emission expected in strict LTE for the  column density and temperature derived from the rotation diagram in Figure 3, but scaled down by a factor of $10^{-4}$. The computed spectra include more than 4000 vibration-rotation lines, each of which has a width of 415 km\,s$^{-1}$. }
\end{center} \end{figure}

\section{Discussion}\label{disc}

To further investigate the physical conditions in the knot we analyse the CO excitation with the non-LTE radiative transfer code {\tt RADEX} \citep{van07}, assuming a uniform emitting region (i.e. constant density, temperature, and abundance).
In order to fit all our rotational lines, as well as the previously reported ro-vibrational emission from the same region \citep{rho09,rho12}, multiple temperature components are required: a $\sim$400~K component, and an additional $\sim$2000~K component, containing about 15\%\ of the CO, to explain the ro-vibrational fluxes (shown in Fig. \ref{fig:AKARI}). This hot component also contributes substantially to the high-J rotational lines. The multiple temperature components are suggestive of the structure of a post-shock cooling zone \citep{bor90}.

The non-LTE analysis yields $N_{CO}\sim 5 \times$10$^{17}$~cm$^{-2}$, comparable to the rotation diagram result. 
The required density of the main collision partner is $\sim$10$^6$~cm$^{-3}$. 
As the knot is expected to be oxygen-rich, with its strong [$\ion{O}{III}$] emission and similarity to optical FMKs, we consider oxygen to be the main collision partner in this knot. However, collisional rates with CO have only been studied with H/H$_2$ and H$_2$O as collision partners. Hence, these rates have been used as proxies for the unknown O + CO collisional rates. 
The derived density differs by an order of magnitude depending on the chosen collision partner. On the other hand, the critical densities of the observed CO lines are $\sim$10$^{6\text{-}7}$~cm$^{-3}$, constraining the gas density to be fairly high.
The {\tt RADEX} analysis indicates that the observed CO lines are optically thin ($\tau < 10^{-2}$).
Overall, the non-LTE analysis results are consistent with the simple rotational diagram analysis.

The high temperature and density of the CO, along with the broad ($\sim$400~km\,s$^{-1}$) lines, suggests that the CO emission originates in a dense knot in the post-shock region of the reverse shock.
In a dense knot, the reverse shock is attenuated due to energy conservation. 
Following \citet{doc10}, a 2000~km\,s$^{-1}$ reverse shock will be slowed to 200~km\,s$^{-1}$ when crossing a knot 100 times denser than the interclump medium. For a 200~km\,s$^{-1}$ shock, the gas temperature at the shock front reaches some 10$^7$~K and any pre-existing molecules, including CO, will be destroyed. 
However, ongoing chemical modeling suggests that, under the observed post-shock conditions, CO will reform on a timescale $t_{CO}$ $\sim$ 100 days, through radiative association reactions (Biscaro \& Cherchneff, in prep.).

Shock attenuation in dense knots also has implications for the survival of SN-produced dust. A slow shock may sputter $<$ 50\% of the dust mass \citep{sil12}, and the warm dense gas we see in the post-shock region is conducive to grain growth. Hence, some SN dust may survive the passage of the reverse shock in dense knots.

A minimum cooling flux, $\Lambda_{min}$, can be estimated from the energy radiated by the observed [$\ion{O}{III}$] and CO lines in our knot. The [$\ion{O}{III}$] 88\,$\mu$m line provides $\sim$1~erg\,cm$^{-2}$\,s$^{-1}$ of cooling, comparable to that of the total rotational CO line emission, giving $\Lambda_{min}$ $\sim$ 2~erg\,cm$^{-2}$\,s$^{-1}$. 
Using a gas column density of 5$\times$10$^{19}$~cm$^{-2}$, as explained below,
we can calculate an approximate gas cooling time ($t_{cool}$=N$_{gas}$$k_B$T/$\Lambda_{min}$) of 70 days. This is comparable to the CO formation timescale $t_{CO} \sim$ 100 days, indicating that a constant heating flux of the order of $\Lambda_{min}$ may be required to maintain the observed large column density of warm CO.

One obvious heat source is the shock front itself, which emits UV photons. 
For a flux of $F_o=n_{o}\times v_{s}$ ($n_o$=oxygen density; $v_s$=shock velocity) oxygen atoms flowing into the shock, the total UV photon flux at the surface of the knot is $30\,F_o$~photons\,cm$^{-2}$\,s$^{-1}$ \citep[Table 11]{bor90}. 
With a typical photon energy of 20~eV, we thus derive a heating flux of $\sim$2~erg\,cm$^{-2}$\,s$^{-1}$, comparable to $\Lambda_{min}$.
However, the penetration depth of the UV photons is only $\sim$3$\times$10$^{17}$~cm$^{-2}$ \citep{bor90}, while the derived $\ion{O}{III}$ column density is $\sim$10$^{19}$~cm$^{-2}$. 
In addition, $N_{CO}\sim 5\times$10$^{17}$~cm$^{-2}$ implies a gas column density of at least 5$\times$10$^{19}$~cm$^{-2}$, as the pre-shock CO abundance in such a knot is expected to be $\sim$10$^{-2}$ \citep{sar13}.
Hence, the UV photons cannot penetrate the full gas column, and additional heating sources must be considered.

The reverse shock traveling through the tenuous inter-knot ejecta creates a hot plasma which will slowly cool through X-rays. Taking the average observed X-ray luminosity over the remnant, $5.5\times 10^{37}$~erg\,s$^{-1}$ \citep{har97}, with a typical photon energy of $\sim$2~keV, we get an X-ray heating flux of 0.2 erg\,cm$^{-2}$\,s$^{-1}$ at the knot. While the keV photons could ionize and heat a much larger column density than the UV photons, $\sim$3$\times$10$^{19}$~cm$^{-2}$, the X-ray heating flux falls short of $\Lambda_{min}$ by an order of magnitude.

Heat conduction from the inter-knot hot, tenuous plasma into the dense knot could provide another heating source. The classical expression for heat conduction by electrons gives $Q = K(T)\,dT/dr$ where $K(T)$ is approximately constant and about $6\times$10$^{-7} \times T^{5/2}$ erg\,s$^{-1}$\,K$^{-7/2}$\,cm$^{-1}$ \citep[p. 448--449]{tie05}. Setting the temperature gradient equal to $T/\delta R$ with the temperature $T$=10$^7$~K and the length scale $\delta R$ given by the gas column density divided by the gas density, $N/n$ = 5$\times$10$^{13}$~cm, we find an energy flux of $\sim$4$\times$10$^{4}$~erg\,cm$^{-2}$\,s$^{-1}$ which will be balanced by mass evaporation from the knot surface. The energy radiated away through the CO lines is then of little relevance for the energy budget. 
For the adopted parameters, the mean free path for electrons ($\sim$10$^{12}$~cm) is small compared with the size scale for the temperature gradient (5$\times$10$^{13}$~cm), as required by the classical heat conduction expression. 
Though the heat flux conducted inwards by electrons may be limited somewhat by magnetic fields, still heat conduction may be key to maintaining the large column density of warm gas.

In conclusion, {\it Herschel} observations of rotational CO lines in a dense knot in Cas A indicate a large column density ($N_{CO}\sim 5\times$10$^{17}$~cm$^{-2}$) of warm (two components at $\sim$400 and 2000~K) and dense (10$^{6\text{-}7}$ cm$^{-3}$) gas in the post-shock region of the reverse shock. The passage of the shock will dissociate any existing molecules and hence the CO has most likely been reformed recently, in the post-shock gas, providing evidence of an active chemistry in the post-reverse-shock region. The observed large column density of warm CO indicates that the cooling through CO (and ionic) lines is balanced by a constant heating flux. The diffuse X-ray flux is insufficient, and the UV photons from the shock front cannot penetrate the full gas column, so heat conduction by electrons may be required to maintain the temperature of the gas.

\begin{acknowledgement}
S.W. and C.B. thank the ESF EuroGENESIS programme for financial support through the CoDustMas network.
\end{acknowledgement}

\end{document}